\documentclass[10pt]{iopart}
\usepackage{graphicx}
\usepackage{textcomp} 
\usepackage{setspace}
\usepackage{hyperref}
\usepackage{amssymb}
\hypersetup{
	colorlinks=true,
	linkcolor=blue,
	anchorcolor=blue,
	citecolor=blue}
\graphicspath{{./figs/}}
\begin{document}

\title[]{Extension of ELM suppression window using n=4 RMPs in EAST}

\author{P. Xie$^{1,2}$, Y. Sun$^{1}$, Q. Ma$^{1,2}$, S. Gu$^{4}$, Y.Q. Liu$^{3}$, M. Jia$^{1}$, A. Loarte$^{5}$, X. Wu$^{1,2}$, Y. Chang$^{1,2}$, T. Jia$^{1,2}$, T. Zhang$^{1}$, Z. Zhou$^{1,2}$, Q. Zang$^{1}$, B. Lyu$^{1}$, S. Fu$^{1,2}$, H. Sheng$^{1,2}$, C. Ye$^{1,2}$, H. Yang$^{1,2}$, H.H. Wang$^{1}$ and EAST Contributors}
\address{$^1$ Institute of Plasma Physics, Hefei Institutes of Physical Science, Chinese Academy of Sciences, Hefei 230031, China}
\address{$^2$ University of Science and Technology of China, Hefei 230026, China}
\address{$^3$ General Atomics, PO Box 85608, San Diego, CA 92186-5608, USA}
\address{$^4$ Oak Ridge Associated Universities, Oak Ridge, TN 37831, USA}
\address{$^5$ ITER Organization, Route de Vinon sur Verdon, 13115 St Paul Lez Durance, France}
\ead{ywsun@ipp.ac.cn}
\vspace{10pt}
\begin{indented}
  \item[]April 2023
\end{indented}

\begin{abstract}
  The $q_{95}$ window for Type-I Edge Localized Modes (ELMs) suppression using $n=4$ even parity Resonant Magnetic Perturbations (RMPs) has been significantly expanded to a range from 3.9 to 4.8, which is demonstrated to be reliable and repeatable in EAST over the last two years.
  This window is significantly wider than the previous one, which is around $q_{95}=3.7\pm0.1$, and is achieved using $n=4$ odd parity RMPs.
  Here, $n$ represents the toroidal mode number of the applied RMPs and $q_{95}$ is the safety factor at the $95\%$ normalized poloidal magnetic flux.
  During ELM suppression, there is only a slight drop in the stored energy ($\leq10\%$).
  The comparison of pedestal density profiles suggests that ELM suppression is achieved when the pedestal gradient is kept lower than a threshold.
  This wide $q_{95}$ window for ELM suppression is consistent with the prediction made by MARS-F modeling prior to the experiment, in which it is located at one of the resonant $q_{95}$ windows for plasma response.
  The Chirikov parameter taking into account plasma response near the pedestal top, which measures the plasma edge stochasticity, significantly increases when $q_{95}$ exceeds 4, mainly due to denser neighboring rational surfaces. 
  Modeling of plasma response by the MARS-F code shows a strong coupling between resonant and non-resonant components across the pedestal region, which is characteristic of the kink-peeling like response observed during RMP-ELM suppression in previous studies on EAST.
  These promising results show the reliability of ELM suppression using the $n=4$ RMPs and expand the physical understanding on ELM suppression mechanism.
\end{abstract}

%
\vspace{5pt}

\section{Introduction}
Serious deterioration of confinement in the plasma pedestal and transient heat pulses on plasma-facing components, such as divertors, caused by Type-I Edge Localized Modes (ELMs), is a critical issue for future fusion devices like ITER \cite{loarte2003characteristics,loarte2014progress,hawryluk2009principal}. 
Resonant magnetic perturbations (RMPs) is an active control method that has been widely and effectively used to control ELMs and the divertor transient heat fluxes in multiple devices \cite{evans2005suppression,evans2015resonant,jeon2012suppression,sun2016nonlinear,suttrop2018experimental}.
Experiments have shown that Type-I ELM suppression with low toroidal mode number RMPs, such as $n = 1$ or 2, has been achieved in wide $q_{95}$ operation windows but is usually accompanied by significant density pump-out and energy confinement degradation, which may prohibit the achievements of high-Q performance in a fusion reactor \cite{evans2015resonant,sun2016nonlinear,suttrop2018experimental,liang2007active,sun2016edge,park2022overview}.
However, experiments of ELM suppression using high-$n$ RMPs have shown big advantages that there is only a slight drop of the energy confinement and minor density pump-out compared to the ELMy H-mode phases and the core tungsten concentration is also reduced \cite{jia2021integrated,sun2021first}.

To ensure that the ITER Project will meet its mission requirements, it is necessary to reduce the ELM energy loss to a acceptable level \cite{hawryluk2009principal} and explore ELM suppression with high-$n$ RMPs across ITER regimes of operation.
In early studies, an ELM suppression window at $q_{95}\approx3.7$ with $\Delta q_{95}\sim0.2$ is achieved in DIII-D using $n=3$ RMPs \cite{evans2005suppression}.
With much lower equivalent input torque, EAST demonstrates the first ELM suppression with $n=4$ odd parity (opposite phases in the upper and lower rows of coil currents) RMPs at $q_{95}$ around 3.7 \cite{sun2021first,gormezano2007chapter}.
However, the $q_{95}$ operation window for ELM suppression is narrow, having a range of approximately $\Delta q_{95}\approx0.2$, which is not sufficient to cover the wide $q_{95}$ range required for different operation phases in the ITER scenario \cite{loarte2020required}.
Importantly, the pre-fusion power operation 1 (PFPO-1) phase of the ITER research plan, which concerns H-mode access and sustainment, operates at a toroidal magnetic field strength of $1.8~\mathrm{T}$, with plasma density ranging from $1.0\times10^{19}$ to $2.0\times10^{19}~\mathrm{m^{-3}}$, and a $q_{95}$ range of 3 to 6 \cite{loarte2021h}.
Meanwhile, the toroidal magnetic field strength, the plasma parameters and the equivalent input torque in EAST are similar to this operation phase in ITER.
Previously, in DIII-D, a wider operation window of ELM suppression was found with $n=3$ RMPs near $q_{95}\sim3.7$ within a range of $\Delta q_{95}>0.7$ at low plasma density \cite{hu2020wide}.
Considering that $n=4$ RMPs configuration is planned in ITER, it is therefore meaningful to further explore a wider ELM suppression window with $n=4$ RMPs in low-torque plasmas in EAST while maintaining good confinement, for the PFPO-1 phase of the ITER research plan.

In this paper, we report further expansion of $q_{95}$ windows for ELM suppression to the ranges $[3.9,~4.1]$ and $[4.2,~4.8]$ with $n=4$ even parity RMPs (with the same phases in the upper and lower rows of coil current) in the EAST experiments in the last two years.
The correlation between ELM suppression windows and edge stochasticity is analyzed based on modeling of plasma response by the linear resistive magnetohydrodynamics (MHD) code MARS-F \cite{liu2010full}.
These results provide insights to understand the mechanism of ELM suppression by RMPs and confirm the use of high $n$ RMPs for ELM suppression with wide operation window to meet the requirements of future ITER operation.

The rest of this paper is structured as follows.
Section \ref{sec:setup} describes the experimental setup in EAST.
In Section \ref{sec:window}, we present experimental observations of ELM suppression with $n=4$ even parity RMPs across a wide range of $q_{95}$.
Section \ref{sec:linear} provides a detailed analysis of the linear modeling of plasma response using MARS-F code, which aids in understanding the optimal RMP coil configuration and the extension of the ELM suppression window.
Finally, we conclude with a summary in Section \ref{sec:discuss}.

\section{Experimental setup}
\label{sec:setup}
\begin{figure}[htbp]
  \centering
  \includegraphics[width=0.35\linewidth]{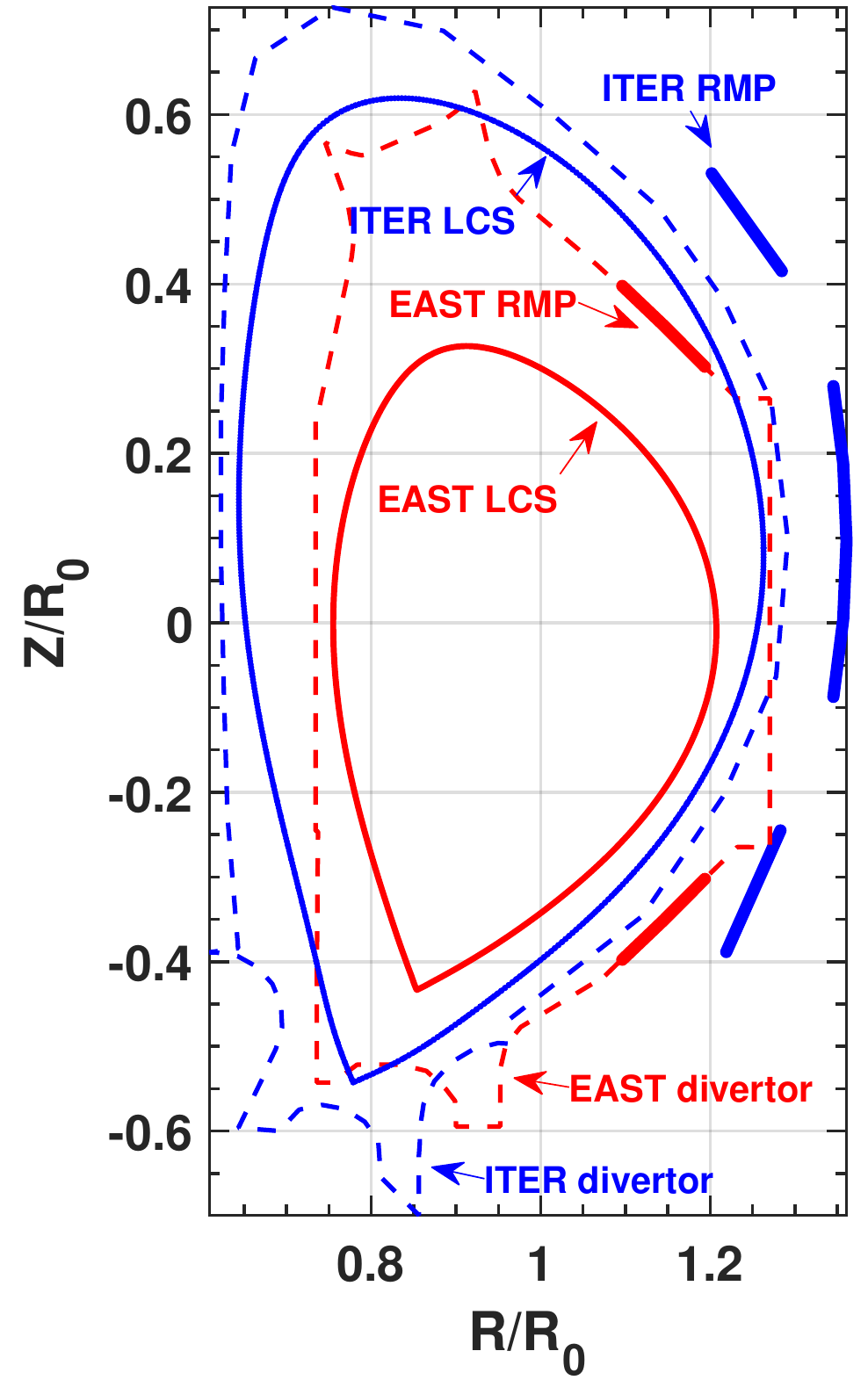}
  \caption{Poloidal cross-section showing the last closed flux surfaces, inner vacuum chamber, and RMP coils in EAST (red) and ITER (blue), normalized to their major radius $R_0$.} 
  \label{fig:EAST-ITER-RMP}
\end{figure}

EAST tokamak has a major radius $R_0$ of $1.8~\mathrm{m}$ and a minor plasma radius of $0.45~\mathrm{m}$ \cite{wan2019recent,wan2020new}.
Figure \ref{fig:EAST-ITER-RMP} shows the poloidal cross-section of the last closed surface of the plasmas, the in-vessel plasma facing components and the RMP coils of both ITER and EAST, which are normalized to their $R_0$ values.
For the EAST RMP system, there are two arrays of coils symmetrically located on the upper (U) and lower (L) parts at the low field side.
Each array consists of eight coils mounted evenly along the toroidal direction, allowing the toroidal mode number of magnetic perturbations up to $n=4$, which is the same as that in ITER \cite{neumeyer2011design}. 
Each coil spans a toroidal angle of $37^\circ$ with four turns, and the maximal total current amplitude amounts to $16~\mathrm{kAt}$.
To generate $n=4$ perturbation fields, the current waveform in adjacent coils of one array is opposite.
The even parity coils configuration means that the RMP coils current waveform is up and down symmetric, and the odd parity represents the opposite-phase current waveform in the upper and lower rows of coils.  

Since 2020, the lower divertor in EAST has been upgraded to an ITER-like tungsten divertor.
In this paper, the plasmas are in the lower single null configuration with lower triangularity of $0.52\sim0.62$. %
The plasma current ranges from $360~\mathrm{kA}$ to $500~\mathrm{kA}$, and $B_\mathrm{t}$ ranges from $1.5~\mathrm{T}$ to $1.7~\mathrm{T}$ in unfavorable $B_\mathrm{t}$ direction ($\nabla B$-drift direction is reverse to the X-point).
The input torque of co-current neutral beam injection (NBI) is around $1~\mathrm{N\cdot m}$, and low hybrid wave (LHW) is employed together for current drive \cite{ding2013experimental}. 

\section{Wide ELM suppression window using $n=4$ even parity RMPs}
\label{sec:window} 
\subsection{ELM suppression with even parity RMPs}
\label{subsec:spectrum}
\begin{figure}[htbp]
  \centering
  \includegraphics[width=0.6\linewidth]{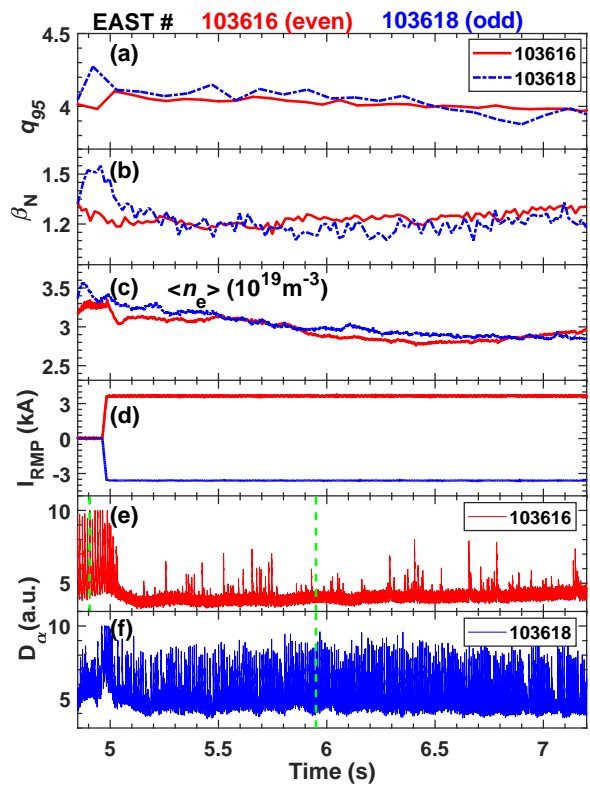}
  \caption{Temporal evolution of (a) $q_{95}$, (b) normalized plasma beta $\beta_N$, (c) central line-averaged densities $<n_e>$, (d) RMP current amplitude, and (e-f) $\mathrm{D}_\alpha$ emission for two discharges 103616 (even) and 103618 (odd) with different coil parity $n=4$ RMPs in EAST.
  }
  \label{fig:q95-4-mit-sup-comp}
\end{figure}

Type-I ELM suppression is achieved in EAST tokamak during the 2021 and 2022 summer experiments using $n=4$ even parity RMPs at $q_{95}\approx4$. 
The RMP-ELM control effects in two discharges with different parity coil configurations are shown in figure \ref{fig:q95-4-mit-sup-comp}. 
In the two discharges, the toroidal magnetic field strength $B_T\sim1.5~\mathrm{T}$, $q_{95}\sim4.0$ and the normalized plasma beta $\beta_N\sim1.2$, as shown in figure \ref{fig:q95-4-mit-sup-comp}(a) and (b).
The heating power includes $1.9~\mathrm{MW}$ co-current NBI and $1.1~\mathrm{MW}$ LHW.
Prior to RMP application, the ELM frequency is approximately $100~\mathrm{Hz}$ as shown in figure \ref{fig:q95-4-mit-sup-comp}(e), the plasma central line-averaged density is around $3.2\times10^{19}~\mathrm{m}^{-3}$ as shown in figure \ref{fig:q95-4-mit-sup-comp}(c), corresponding to $47\%$ of the Greenwald density $n_{\mathrm{GW}}$, and the normalized electron collisionality near the pedestal is around $\nu_{*e,ped}\sim0.4$.
With even parity RMPs application, ELM suppression is observed in discharge 103616 (figure \ref{fig:q95-4-mit-sup-comp}(e)).
In contrast, only mitigation is achieved with odd parity RMPs with the same coils current in discharge 103618, during which the ELM frequency is increased to $160~\mathrm{Hz}$ (figure \ref{fig:q95-4-mit-sup-comp}(f)). 
Although the influences of different parity coil configurations on ELM are distinct, the drop in particle and energy confinement is slight and around $10\%$ in both discharges.
Due to the impurities accumulation in discharge 103618, the bursts of ELMs are irregular before $4.95\mathrm{s}$, but return to a regular pattern before the RMP application.
These experiments demonstrate that coil configuration plays a crucial role in ELM control effectiveness, and ELM suppression can be achieved in EAST plasmas with $q_{95}\sim4.0$ using $n=4$ even parity RMPs.

\begin{figure}[htbp]
  \centering
  \includegraphics[width=0.6\linewidth]{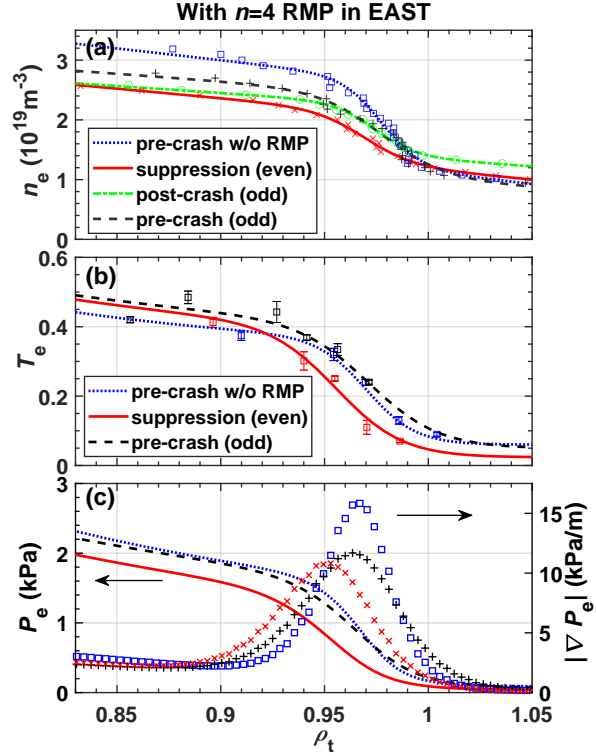}
  \caption{(a) The density pedestals in the $\sim80\%$ of the ELM cycle ($4905~\mathrm{ms}$, blue dotted line) without RMPs and ELM suppression state ($5950~\mathrm{ms}$, red solid line) in discharge 103616, as well as the $\sim80\%$ ($5957~\mathrm{ms}$, black dashed line) and $\sim20\%$ ($5959~\mathrm{ms}$, green dot-dashed line) of the ELM cycle in discharge 103618, measured at $Z=0.03~\mathrm{m}$ with the reflectometry in EAST.
  (b) The electron temperature pedestal in the $\sim90\%$ of the ELM cycle ($4916~\mathrm{ms}$, blue dotted line) without RMPs and ELM suppression state at $5916~\mathrm{ms}$ (red solid line) in discharge 103616, as well as the $\sim80\%$ of the ELM cycle ($5916~\mathrm{ms}$, black dashed line) in discharge 103618, measured at $R=1.90~\mathrm{m}$ with the Thomson Scattering in EAST.
  (c) The electron pressure pedestal given by the product of densities and temperature.
  The connecting lines represent the fitted tanh function.
  }
  \label{fig:ne-pedestal}
\end{figure}

To investigate how pedestal stability is influenced by different parity RMPs, the pedestal profiles of electron density and temperature in the two discharges, measured by reflectometry and Thomson Scattering, respectively \cite{qu2015q,zang2016characteristics}, are compared.
Figure \ref{fig:ne-pedestal}(a) displays the density profiles in the $\sim80\%$ of the ELM cycle (pre-crash) without RMPs and at the ELM suppression moment (suppression) with even parity RMPs in discharge 103616, as well as the $80\%$ (pre-crash) and $20\%$ (post-crash) of the ELM cycle with odd parity RMPs in discharge 103618.
The fitted profiles with a modified tanh function are represented by lines with the same color as the corresponding data.
Prior to the ELM crash (pre-crash) without RMPs, the density pedestal is the highest and steepest.
In the mitigation case, RMPs cause a $10\%$ decrease in density at the top pre-crash and reduce its gradient.
The density pedestal drops a further $\sim5\%$ after the ELM crash (post-crash), and there is a significant increase in density in the SOL region, resulting in a reduction of the pedestal gradient.
In the ELM suppression case, the density pedestal at the top is similar to the post-ELM crash one, and the pedestal width is slightly wider.

Figure \ref{fig:ne-pedestal}(b) displays the temperature profiles in the $\sim90\%$ of the ELM cycle (pre-crash) without RMPs and the ELM suppression moment (suppression) with even parity RMPs, and the $\sim80\%$ (pre-crash) of the ELM cycle with odd parity RMPs are compared.
The temporal resolution of Thomson Scattering in this experiment is not high enough to capture the moment post-ELM crash during ELM mitigation.
Regarding electron temperature, the case pre-crash during ELM mitigation is slightly higher than that without RMPs, which is consistent with the observations in Ref. \cite{evans2008rmp}.
The pedestal during ELM suppression has a similar level at the top as the mitigation case, but moves inward around $2\%$ in $\rho_t$ ($\sim1~\mathrm{cm}$) compared to the case without RMPs, suggesting the effect of RMP-induced 3D displacement.
The electron pressure profiles, shown as figure \ref{fig:ne-pedestal}(c), are given by the product of density and temperature fitted profiles.
The gradient of the case pre-crash during ELM mitigation is lower than that without RMPs, mainly contributed by its lower density pedestal.
The electron pressure gradient further slightly drops during ELM suppression.
And its inward shift is due to the electron temperature pedestal.
The comparison of changes in the pedestal profiles suggests that ELM suppression is achieved when the pedestal gradient is kept lower than a threshold. 

\subsection{Extension of ELM suppression window}
\label{subsec:extended}
\begin{figure}[htbp]
  \centering
  \includegraphics[width=0.7\linewidth]{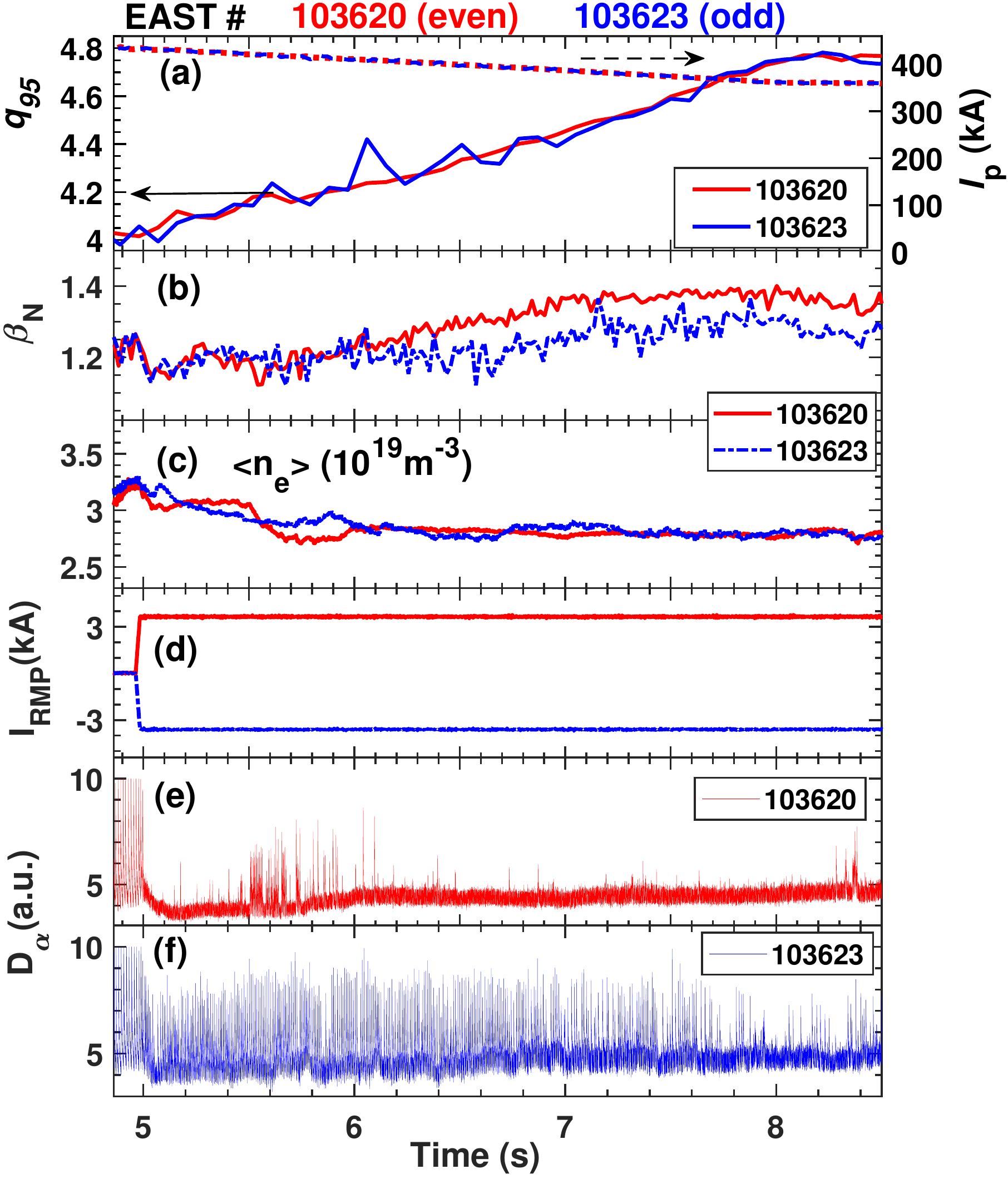}
  \caption{Temporal evolution of (a) $q_{95}$ and plasma current $I_\mathrm{p}$, (b) normalized plasma beta $\beta_N$, (c) central line-averaged densities $<n_e>$, (d) RMP current amplitude, and (e-f) $\mathrm{D}_\alpha$ emission for two discharges 103620 (even) and 103623 (odd) with different coil parity $n=4$ RMPs in EAST.
  }
  \label{fig:q95-4-103620}
\end{figure}

ELM suppression has been extended to a higher $q_{95}$ above 4 with $n=4$ even parity RMPs in EAST, shown as figure \ref{fig:q95-4-103620}.
The plasmas parameters before RMP application and the heating condition are the same as the experiments mentioned in section \ref{subsec:spectrum}.
Heating power is kept the same during the ramp up of $q_{95}$ while plasma current is slowly ramp down from $430~\mathrm{kA}$ to $360~\mathrm{kA}$, i.e, $15\%$ decreasing in $3~\mathrm{s}$ or about $18~\mathrm{kA/s}$, which is much longer than current diffusion time around $0.5~\mathrm{s}$ in EAST.
Figure \ref{fig:q95-4-103620}(b) shows that the normalized beta $\beta_N$ gradually increase from 1.1 to 1.4, mainly due to $q_{95}$ increase, though the stored energy drops around $10\%$. 
In discharge 103620 (figure \ref{fig:q95-4-103620}(e)), ELM suppression sustains with even parity RMPs when $q_{95}$ slowly ramps up from 4.0 to 4.8 over a three-second interval except a short period at $q_{95}\approx4.1\sim4.2$. 
ELM mitigation is observed in most of the period in another similar discharge 103623 but odd parity of $n=4$ RMPs in EAST (figure \ref{fig:q95-4-103620}(f)).
The ELM frequency increases about twice as much compared to that before RMP application.
ELM suppression is only achieved in this odd parity case near $t=8~\mathrm{s}$, when $q_{95}\sim4.75$.
During RMP application, the plasma central line-averaged density (figure \ref{fig:q95-4-103620}(c)) slightly drops around $10\%$, and no obvious drop of the central ion and electron temperature occurs.
The experiments demonstrate a much wider $q_{95}$ window in $[4.2,~4.8]$ for ELM suppression with $n=4$ even parity RMPs in addition to the ELM suppression with $q_{95}\sim4.0$.

\subsection{$n=4$ RMP ELM suppression windows in EAST}
\label{sec:domain}
\begin{figure}[htbp]
  \centering
  \includegraphics[width=0.6\linewidth]{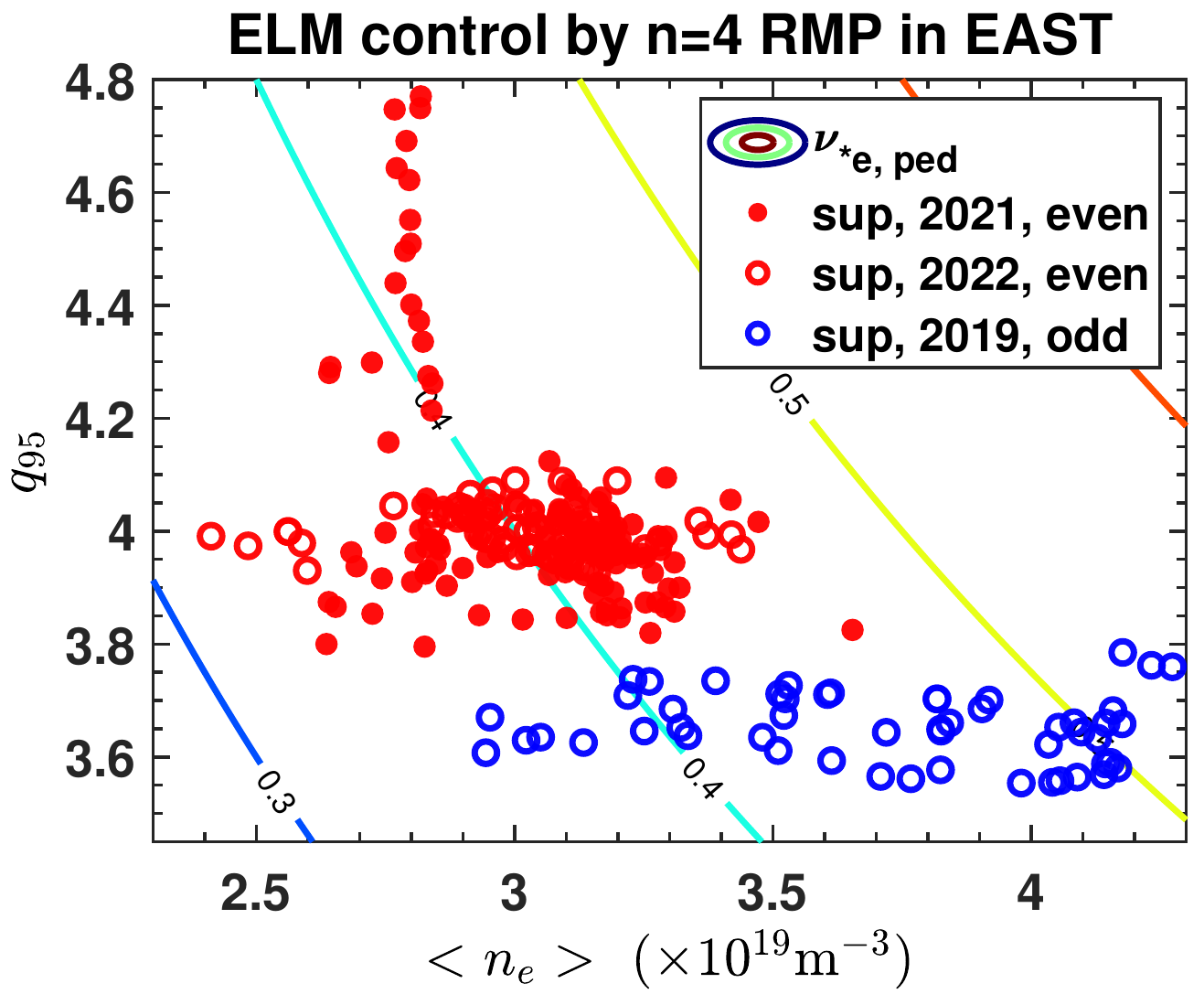}
  \caption{A summary of ELM suppression windows achieved with $n=4$ RMPs in EAST.
    Red filled circles and hollow blue circles represent ELM suppression with even parity RMPs in 2021 and 2022 experiments separately.
    Blue circles are that with odd parity RMPs in 2019 experiments from Ref.\cite{sun2021first} figure 6.
    The Contour line of pedestal collisionality is also plotted on the same graph, with the assumption of $T_\mathrm{e,~ped}=0.6~\mathrm{keV}$, $n_\mathrm{e,~ped}=0.75\times\left\langle n_e\right\rangle$, and $Z_\mathrm{eff}=1$.
  }
  \label{fig:n4-19-21A}
\end{figure}

The ELM suppression window with $n=4$ even parity RMPs in EAST is reliable and spans a wide $q_{95}$ range, supported by experimental data from the 2021 and 2022 campaigns.
A summary of ELM suppression windows with $n=4$ RMPs is shown in figure \ref{fig:n4-19-21A}, consisting of hundreds of time slices from over eighty stable ELMy H-mode discharges.
Here, red filled circles and hollow circles represent the moments of ELM suppression with even parity RMPs in 2021 and 2022 experiments separately, and hollow circles correspond to those with odd parity RMPs.
For ELM suppression with even parity RMPs, $q_{95}$ window crosses widely from 3.9 to 4.8, and the plasma central line-averaged density ranges from $2.5\times10^{19}$ to $3.5\times10^{19}~\mathrm{m^{-3}}$, corresponding to $40\%\sim53\%~n_{\mathrm{GW}}$.
The experimental data from 2021 and 2022 campaigns overlap, indicating the experimental repeatability of the ELM suppression window.
Additionally, the statistical average drop in plasma density during ELM suppression is around $10\%$ compared to the ELMy H-mode phases, while the average drop in stored energy and confinement factor $\mathrm{H}_{98,y2}$ is more slight and less than $5\%$.

The comparison between the ELM suppression windows obtained with $n=4$ odd and even parity RMP coil configurations reveals obvious differences in $q_{95}$ and the plasma central line-averaged density.
With the approximate pedestal collisionality of $\nu_{*e}\sim0.4$, the ELM suppression window obtained with the $n=4$ odd parity RMPs is located at $q_{95}\sim3.7$ with a range of $\Delta q_{95}\sim0.2$, in which the plasma line-averaged density is $15\%$ higher than that with even parity.
There is a distinct gap of $\Delta q_{95}\approx1/4$ between the two ELM suppression windows, which is the periodicity that results in one more rational surface appearing when $q_{95}+1/4$.
This suggests a potential important interaction between the plasma response induced by RMPs and a special local place of pedestal.
Moreover, a narrow ELM suppression window at $q_{95}\approx3.7$ is also observed in DIII-D using $n=3$ RMPs \cite{evans2005suppression}, which implies the similar effect of high $n$ magnetic perturbations in the pedestal.
The statistical summary demonstrates that the $q_{95}$ window for ELM suppression using $n=4$ even parity RMPs is much wider than before.
The confinement can be well maintained during ELM suppression within this operation window.

\section{Understanding the wide ELM suppression window}
\label{sec:linear}
\subsection{Numerical setting}
The MARS-F code calculates the plasma response to external magnetic perturbations by solving the linear resistive MHD equations.
By simulating the plasma response to magnetic perturbations, MARS-F provides insights into the mechanisms underlying ELM suppression, including their effects on pedestal transport and peeling-ballooning mode growth rate \cite{evans2008rmp,liu2016elm}.
The code takes into account plasma resistivity and rotation, and has been validated by numerous experiments \cite{lanctot2010validation,wang2015three}.
The linear MHD equations solved in the MARS-F code can be written as \cite{liu2010full,liu2011modelling}
\begin{eqnarray}
  &&i\left(\Omega_\mathrm{RMP}+n \Omega\right) \xi = v +( \xi \cdot \nabla \Omega) R \hat{\phi},    \label{eqn:mars1}                                                                                                                                                                                                                             \\
  &&i \rho\left(\Omega_\mathrm{RMP}+n \Omega\right) v =-\nabla p+ j \times B + J \times b                          -\rho[2 \Omega \hat{ Z } \times v      \nonumber\\
    &&\qquad+( v \cdot \nabla \Omega) R \hat{\phi}]-\rho \kappa_{\|}\left|k_{\|} v_{ th , i }\right|\left[ v +( \xi \cdot \nabla) V _{0}\right]_{\|}, \label{eqn:mars2} \\
  &&i\left(\Omega_\mathrm{RMP}+n \Omega\right) b=\nabla \times( v \times B )+( b \cdot \nabla \Omega) R \hat{\phi}  -\nabla \times(\eta j ), \label{eqn:mars3}                                                                                                                                                                                    \\
  &&i\left(\Omega_\mathrm{RMP}+n \Omega\right) p=- v \cdot \nabla P-\Gamma P \nabla \cdot v, \label{eqn:mars4}                                                                                                                                                                                                                                    \\
  &&j=\nabla \times b, \label{eqn:mars5}
\end{eqnarray}
where $V_0=R\Omega\hat{\phi}$, and the variables $\eta$, $\xi$, $v$, $b$, $j$, $p$, $\Omega$ represent the plasma resistivity, plasma displacement, perturbed velocity, magnetic field, current, pressure, and toroidal angular frequency, respectively.

The input equilibrium used in the modeling is reconstructed with k-EFIT \cite{lao1990equilibrium,li2013kinetic}, at $5000~\mathrm{ms}$ in EAST discharge 103041 with parameters of $n_e\sim3.1\times10^{19}~\mathrm{m^{-3}}$, $\beta_N\sim1.3$ similar to those in discharge 103016.
The toroidal magnetic field $B_t=1.5~\mathrm{T}$, plasma current $I_p=420~\mathrm{kA}$, and $q_{95}=4.0$.
The plasma heating condition is the same as the experiments mentioned in section \ref{subsec:spectrum}, where the NBI torque is around $0.9~\mathrm{N\cdot m}$ \cite{pankin2004tokamak}, equivalent to the NBI torque of $28~\mathrm{N\cdot m}$ in ITER, using same calculation method in Ref. \cite{sun2021first}. 
It is noteworthy that the initial modeling was done prior to the experiment using an equilibrium in a similar previous discharge to guide the optimization of RMPs coil configurations and $q_{95}$ operation window.

\begin{figure}[htbp]
  \centering
  \includegraphics[width=0.65\linewidth]{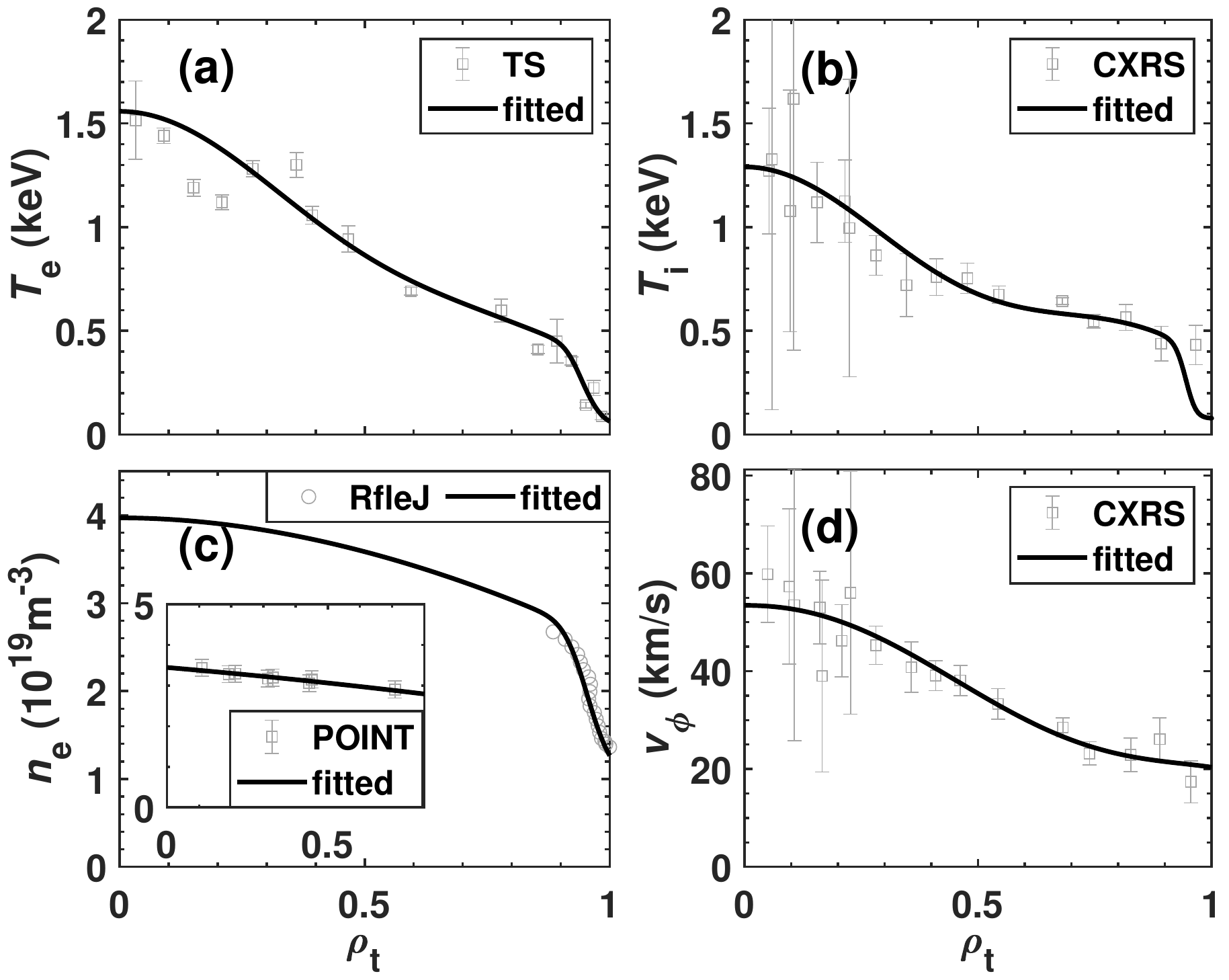}
  \caption{Reference kinetic profiles of electron temperature (a), ion temperature (b), plasma density (c), and toroidal rotation (d) in EAST discharge 103041 used in the modeling.}
  \label{fig:prof103041t5}
\end{figure}

Here, the kinetic profiles in EAST discharge 103041 used in the modeling are shown in figure \ref{fig:prof103041t5}. 
The normalized radius is represented as $\rho_t=\sqrt{\psi_t}$, where $\psi_t$ is the normalized toroidal magnetic flux.
Thomson Scattering (TS) gives the profile of electron temperature (figure \ref{fig:prof103041t5}(a)) \cite{zang2016characteristics}, and the Charge eXchange Recombination Spectroscopy (CXRS) \cite{li2014development} gives the profiles of ion temperature (figure \ref{fig:prof103041t5}(b)) and toroidal rotation (figure \ref{fig:prof103041t5}(d)). 
The electron density profile (figure \ref{fig:prof103041t5}(c)) is obtained with Reflectometry \cite{qu2015q} and together with the constraints on line integrated density measured by POlarimeter-INTerferometer (POINT) \cite{liu2014faraday} in EAST. 

\subsection{Understanding the extension of ELM suppression window}
To understand the wide $q_{95}$ window in $[4.2,~4.8]$ for ELM suppression window with $n=4$ even parity RMPs, $q_{95}$ dependence of the plasma response is modelled using the MARS-F code.
To achieve ELM suppression with RMPs, two key criteria have been identified: the normal plasma displacement near the X-point \cite{liu2016elm} and the Chirikov parameter \cite{chirikov1979universal}.
The normal plasma displacement near the X-point, which is normal to the equilibrium magnetic field, is an important parameter that can help to identify the type of plasma response to the RMP fields and provide insight into the mode structure of the plasma response \cite{liu2016elm}.
The Chirikov parameter is a quantitative measure of edge stochasticity and island overlapping condition which can affect the pedestal structure and stability \cite{evans2004suppression,sun2015modeling}, and can be expressed as
\begin{equation}
  \sigma_{12}=\frac{w_{m_1,n_1}+w_{m_2,n_2}}{2\left|\rho_2-\rho_1\right|}
\end{equation}
where $\sigma_{12}$ is the Chirikov parameter determined by two neighboring islands located at $\rho_1$ and $\rho_2$, and $w_{m,n}$ is the width of the local magnetic island.
The width of the triggered magnetic islands $w$ \cite{xie2021plasma} and the Fourier components of the normalized radial magnetic perturbation field taking into account of plasma response $b_{mn}$ \cite{becoulet2008numerical} can be expressed as
\begin{eqnarray}
  w_{mn}&=4\sqrt{\left|\frac{\rho b_{mn}}{nS}\right|}_{q=q_s}\\
  b_{mn}&=\frac{B^\rho}{B^\zeta}\mathrm{e}^{-i(m\theta-n\zeta)}
\end{eqnarray}
where $m$ and $n$ are the poloidal and toroidal mode numbers respectively, $S=\rho q^\prime/q$ is the global magnetic shear, $B^\rho$ represents radial components of magnetic perturbations with plasma response taken into account, and $B^\zeta$ is the toroidal equilibrium field.

\begin{figure}[htbp]
  \centering
  \includegraphics[width=0.5\linewidth]{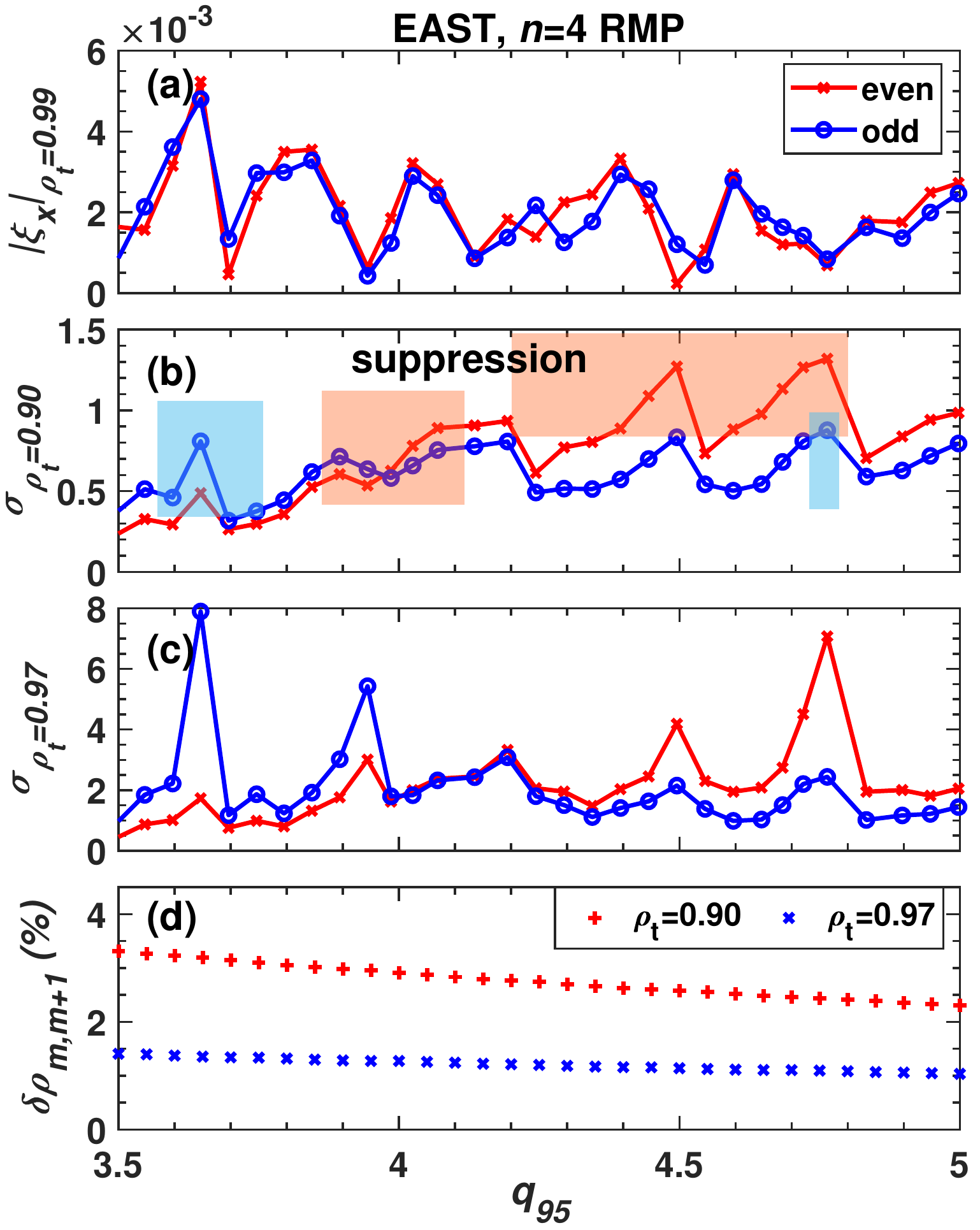}
  \caption{MARS-F plasma response modeling results on $q_{95}$ dependence of boundary normal plasma displacement near the X-point (a) and Chirikov parameter near the pedestal top (b, $\rho_t=0.90$) and bottom (c, $\rho_t=0.97$) for odd (blue circles) and even (red pluses) parity $n=4$ RMP coil configurations, and distances between two neighboring rational surfaces near $\rho_t=0.9$ (d).
    The shaded area indicates the ELM suppression windows confirmed in the experiments in EAST, as figure \ref{fig:q95-4-103620}, \ref{fig:n4-19-21A} shown.
  }
  \label{fig:q-scan2d-103041-sigma}
\end{figure}

Modeling results on $q_{95}$ dependence of plasma response using the MARS-F code \cite{liu2010full} are presented in figure \ref{fig:q-scan2d-103041-sigma}, which shows the normal plasma displacement near the X-point (figure \ref{fig:q-scan2d-103041-sigma}(a)) and Chirikov parameter near the pedestal top (figure \ref{fig:q-scan2d-103041-sigma}(b)) and bottom (figure \ref{fig:q-scan2d-103041-sigma}(c)) for even and odd parity RMPs at different values of $q_{95}$.
Here, a reference equilibrium with $q_{95}=4.0$, as mentioned above, is used for generating a series of equilibrium with different $q_{95}$ by scaling plasma current using CHEASE \cite{lutjens1996chease}.
A fixed $\beta_N\sim1.2$ is used in the equilibrium scan to exclude the beta effect on plasma response \cite{liu2010full}. 

Both criteria show similar location of resonant windows and periodicity in $q_{95}$.
The normal plasma displacement near the X-point for the different parity RMPs is approximate, and does not show a clear tendency for an ELM suppression window with even parity RMPs.
However, the Chirikov parameter near the pedestal top becomes larger as $q_{95}$ exceeds 4, which would be further larger considering the beta effect \cite{liu2010full}, indicating a strong relationship with the wide ELM suppression of $q_{95}$ in $[4.2,~4.8]$ for even parity RMPs.
The Chirikov parameter for odd parity also shows resonant windows at $q_{95}\sim4.5$, but the level is not obviously higher than that at $q_{95}\sim3.9$, which may not be sufficient for achieving wide ELM suppression window.
The shaded area indicates the ELM suppression windows of $q_{95}$ in $[3.9,~4.1]$ and $[4.2,~4.8]$ for even parity RMPs, and the windows of $q_{95}$ in [3.5,~3.7] and $4.75$ for odd parity RMPs  (see Fig. \ref{fig:q95-4-mit-sup-comp}(e), Fig. \ref{fig:q95-4-103620}(e) and (f), and Fig. \ref{fig:n4-19-21A}).
The Chirikov parameter near the pedestal bottom shows similar resonant windows as those in the pedestal top except that of $q_{95}\sim4.0$ for even parity RMPs, but its improvement as $q_{95}$ increase is not that obvious.
The main reason is the distances between two neighboring rational surfaces near the pedestal top becomes narrower as $q_{95}$ increases as shown in figure \ref{fig:q-scan2d-103041-sigma}(d), leading to denser island overlapping.
It should be noted that this modeling was initially done before the experiment to guide the search for new $q_{95}$ suppression windows and the selection of coil parity for $n=4$ RMPs.
The Chirikov parameter near the pedestal from the linear plasma response modeling top provides a useful criterion for assessing ELM suppression. 
It could be applied across devices and used to predict ELM suppression requirements in ITER.

\subsection{Understanding the optimal coil parity for $n=4$ RMPs}
To investigate the mechanism on ELM suppression, the spectra of response field for different parity RMPs are analyzed.
ELM suppression is a nonlinear process and related to the penetration of magnetic perturbations at the pedestal top \cite{sun2016nonlinear,snyder2012eped,hu2020wide}, during which the magnetic islands formed by resonant Fourier components of the plasma response plays a crucial role \cite{wade2015advances,nazikian2015pedestal}.

\begin{figure}[htbp]
  \centering
  \includegraphics[width=0.5\linewidth]{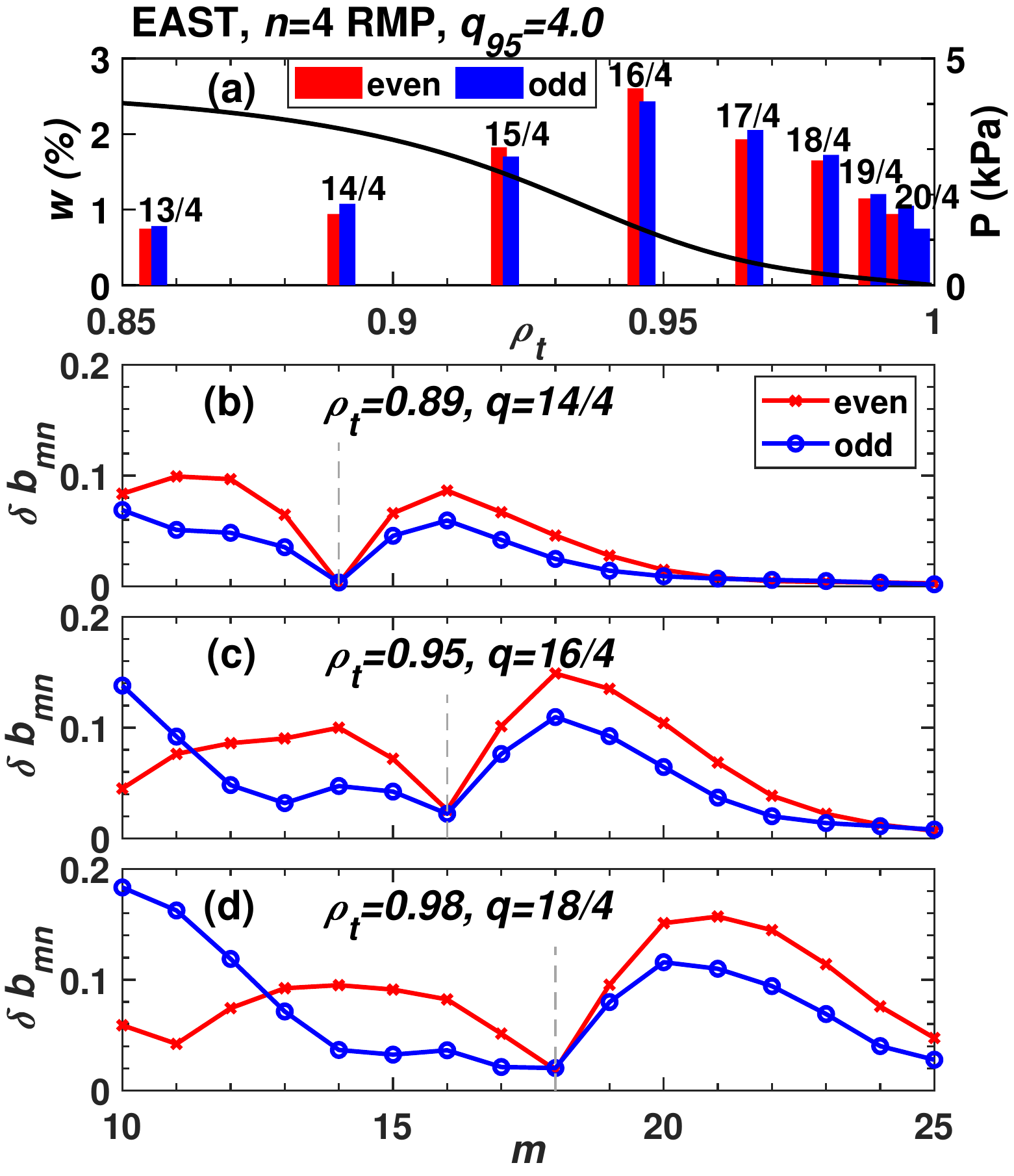}
  \caption{modeling results of plasma response to $n=4$ even (red) and odd (blue) parity RMPs using the MARS-F code, for the equilibrium with $q_{95}=4.0$.
    (a) the radial distribution of magnetic islands induced by the RMPs, where the bars represent the width of the magnetic islands.
    (b-d) the Fourier components of plasma response to $n=4$ RMPs at different rational surfaces, where gray vertical lines indicate the poloidal mode number of resonant harmonics.
  }
  \label{fig:spectrum-even-odd}
\end{figure}

\begin{figure}[htbp]
  \centering
  \includegraphics[width=0.5\linewidth]{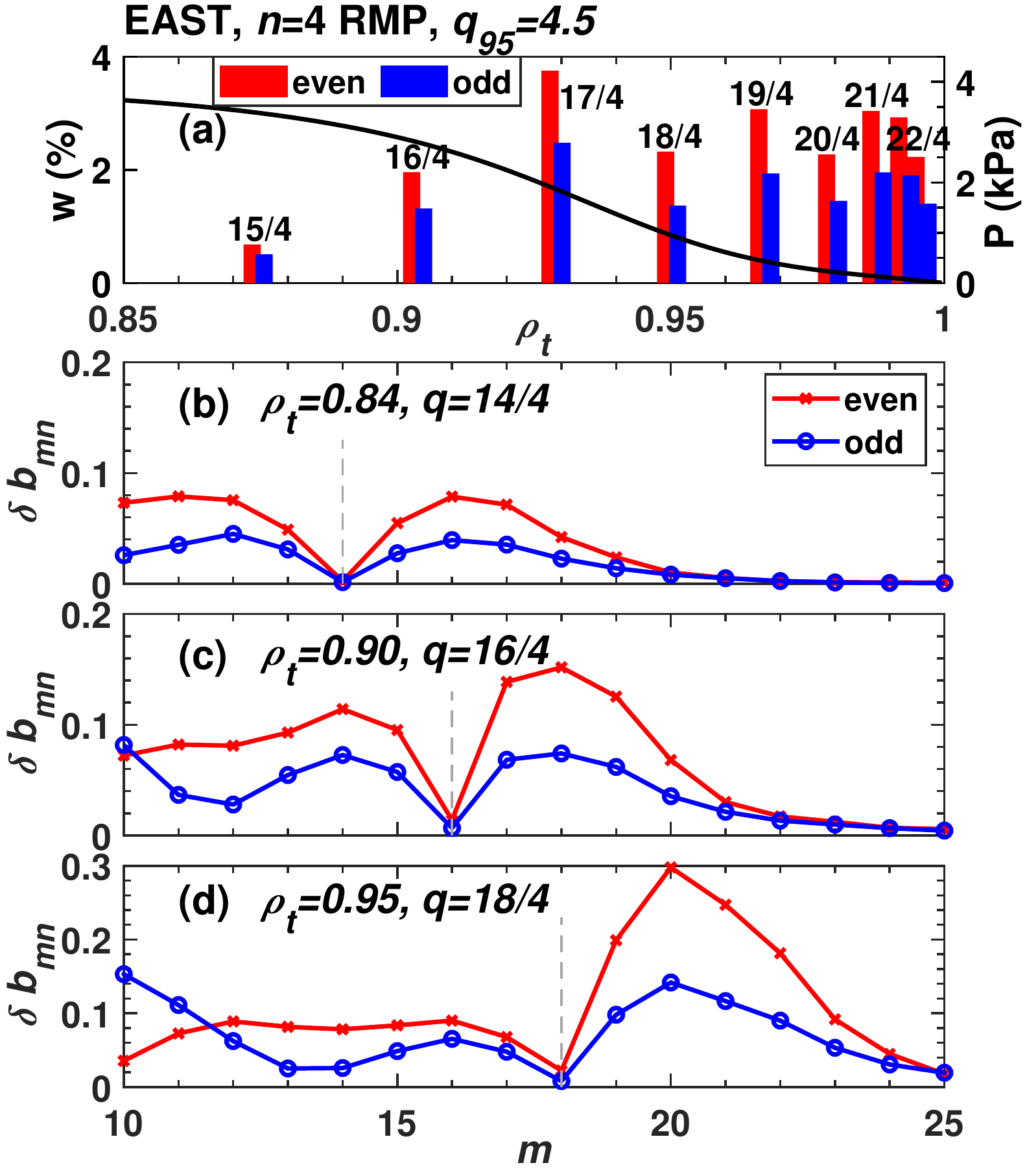}
  \caption{modeling results of plasma response to $n=4$ even (red) and odd (blue) parity RMPs using the MARS-F code, for the equilibrium with $q_{95}=4.5$.
    (a) the radial distribution of magnetic islands induced by the RMPs.
    (b-d) the Fourier components of plasma response to $n=4$ RMPs at different rational surfaces.
  }
  \label{fig:spectrum-even-odd-q450}
\end{figure}

Figure \ref{fig:spectrum-even-odd}(a) presents the width of the magnetic islands for different parity RMPs.
The islands induced by even parity RMPs is only slightly wider than that by odd parity RMPs near the pedestal top.
Figure \ref{fig:spectrum-even-odd}(b$\sim$d) shows Fourier components of plasma response for even and odd parity RMPs at different rational surfaces.
The gray vertical lines indicate the poloidal mode number of resonant harmonics, of which amplitude is relatively low compared to non-resonant harmonics because of plasma screening effect \cite{liu2011modelling}.
For non-resonant components near the rational surfaces, the difference in amplitude between even and odd parity is up to a factor of 3.
The characteristic for even parity RMPs is known as the kink-peeling like response \cite{liu2011modelling} and has been observed in previous ELM suppression experiments on EAST \cite{sun2016edge,xie2021plasma}.
Nonlinear effects like Neoclassical Toroidal Viscosity (NTV) induced by the non-resonant magnetic field has strong effect on plasma toroidal flow and field penetration \cite{sun2010neoclassical,ye2023effect}.
The result suggests that the nonlinear effect on field penetration may play an important role in high $n$ RMPs ELM control.
Compared to the equilibrium with $q_{95}=4.0$, the modeling plasma response for the equilibrium with $q_{95}=4.5$ show (figure \ref{fig:spectrum-even-odd-q450}), the resonant components of even parity RMPs become much more dominant and the ratio of the amplitude of non-resonant components  between different RMPs also enlarges. 
In general, stronger plasma response in both resonant and non-resonant harmonics contributes to ELM suppression with even parity RMPs.

\subsection{Mode coupling effect in plasma response}
\begin{figure}[htbp]
  \centering
  \includegraphics[width=0.5\linewidth]{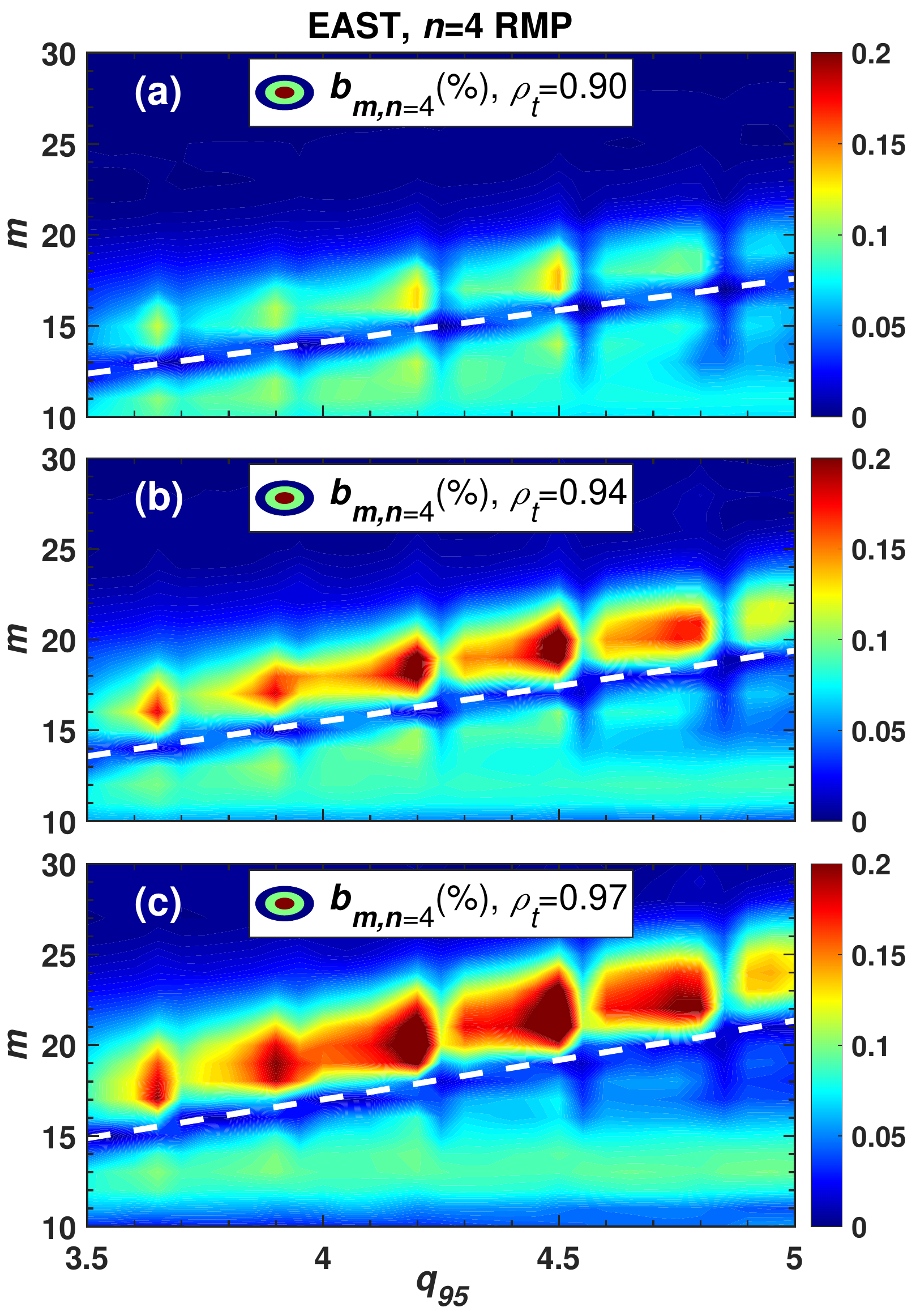}
  \caption{MARS-F plasma response modeling results on $q_{95}$ dependence of RMP spectrum on poloidal mode number near $\rho_t=0.9$ (a), $0.94$ (b), $0.97$ (c) for even parity coil configuration with dashed line indicates $m = nq$.
  }
  \label{fig:mode-structures}
\end{figure}

The mode structures of plasma response to even parity RMPs are further analyzed and $q_{95}$ dependence of the Fourier harmonics across the pedestal are presented in figure \ref{fig:mode-structures}, where the dashed lines indicate $m=nq$.
The presence of multiple resonant windows in adjacent non-resonant harmonics as those in resonant harmonics (figure \ref{fig:q-scan2d-103041-sigma} (b)) suggests a strong poloidal mode coupling in the mode structure, where the amplified kink-peeling response could drive the resonant harmonics \cite{haskey2014linear,ryan2015toroidal}.
The coupling effect is significant across the pedestal, which is a characteristic of kink-peeling response in the EAST ELM suppression plasmas \cite{sun2016edge,sun2021first}. 

\begin{figure}[htbp]
  \centering
  \includegraphics[width=0.5\linewidth]{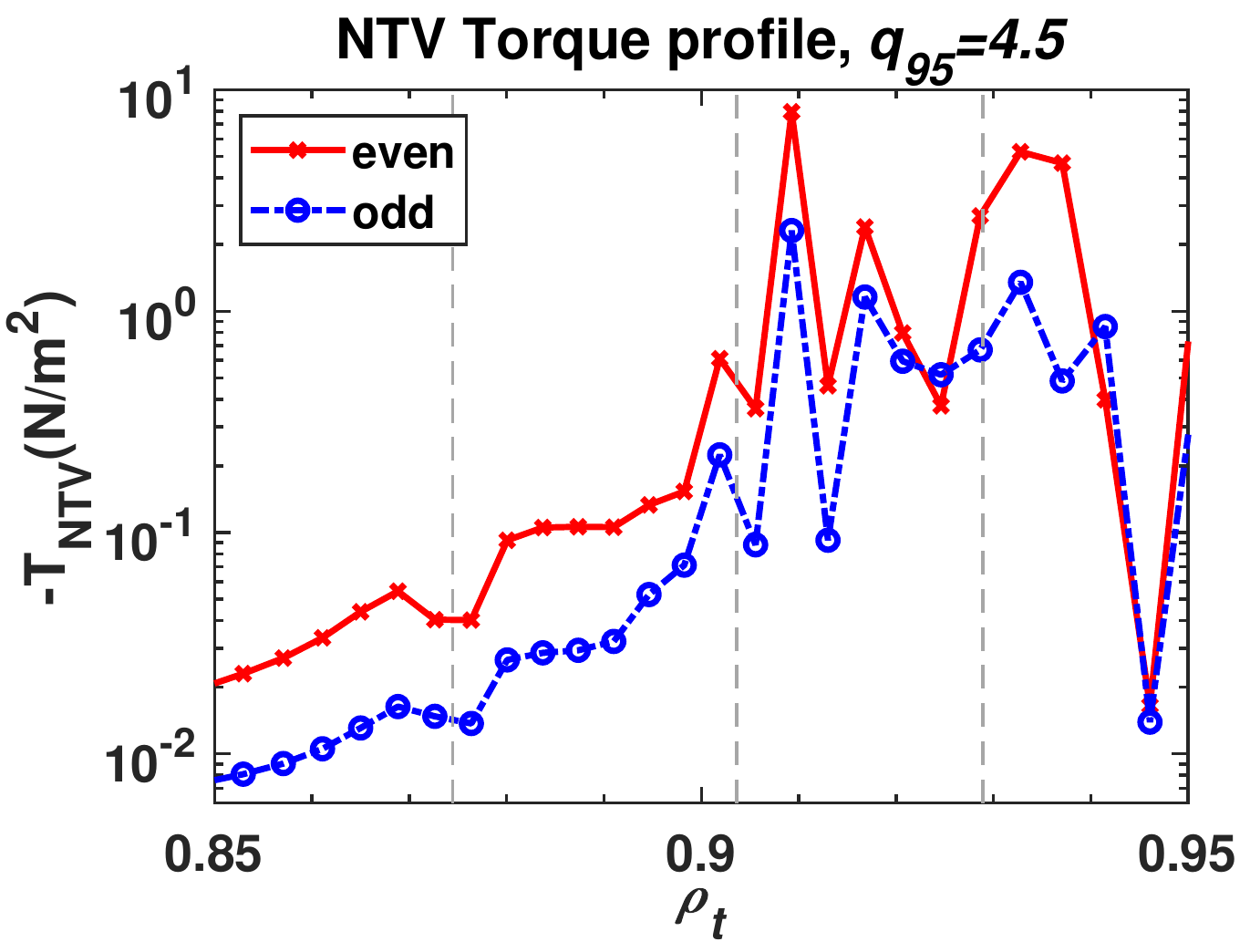}
  \caption{The radial profiles of the NTV torque density for even parity RMPs (red crosses) and odd parity RMPs (blue circles) for the equilibrium with $q_{95}=4.5$.
    The dashed grey lines indicate the location of rational surfaces.
  }
  \label{fig:even-odd-semi-ntv}
\end{figure}

The coupling between resonant and non-resonant harmonics can increase the NTV torque. 
Figure \ref{fig:even-odd-semi-ntv} shows the profiles of the NTV torque density in the plasma edge for even and odd parity RMPs for the equilibrium with $q_{95}=4.5$ by NTVTOK code modeling \cite{sun2011modelling}.
The torque density of even parity RMPs is overall larger than that of odd parity RMPs, which suggests stronger effects on edge plasma rotation and perturbation fields penetration caused by plasma response.
In this case, the NTV torque contributed by ions is comparable to that of electrons, and their directions are opposite.
The results show that even parity RMPs induces larger NTV torque in the plasma edge with denser resonant and non-resonant harmonics, which could significantly impact plasma toroidal flow that is closely related to the penetration of perturbation fields \cite{lee2016effects}.

\section{Summary}
\label{sec:discuss}
The paper presents the extension of ELM suppression $q_{95}$ window with $n=4$ even parity RMPs in recent experiments in EAST and provides the modeling results with good experimental consistency on $q_{95}$ dependence of plasma response using the MARS-F code.
The reliable ELM suppression windows of $q_{95}$ in $[3.9,~4.1]$ and $[4.2,~4.8]$ are much wider compared to the previous $q_{95}$ window of $\Delta q_{95}\sim0.2$ with odd parity RMPs in EAST \cite{sun2021first}.
The advantages of using $n=4$ RMPs are reflected in a much smaller drop ($\leq10\%$) in the plasma stored energy and density when ELM suppression is achieved, compared to $n = 2$ ($\gtrsim20\%$) in EAST \cite{jia2021integrated}. 
The density pedestal at ELM suppression has the lowest gradient among different ELM states suggesting that ELM suppression is achieved when the pedestal gradient is kept lower than a threshold. 
The plasma response from MARS-F linear modeling shows a strong correlation between the Chirikov parameter near the pedestal top and ELM suppression in the range of $q_{95}$ between 4.2 and 4.8 for even parity RMPs, while there is no obvious tendency observed in the boundary normal plasma displacement in this case.
The Chirikov parameter increases as $q_{95}$ exceeds 4 and more edge rational surfaces appears, mainly due to the closer spacing between two neighboring rational surfaces.
The detailed comparison of resonant and non-resonant in plasma response from two equilibria indicates the stronger plasma response contributes to ELM suppression with $n=4$ RMPs.
The presence of multiple resonant windows in both resonant and non-resonant harmonics of plasma response further indicates a strong poloidal mode coupling due to toroidal effect in the mode structure, which is characteristic of the kink-peeling like response observed during EAST RMP-ELM suppression.
The NTV torque for even parity is overall larger than that for odd parity, which could have strong effect on plasma toroidal flow and thus nonlinear perturbation fields penetration \cite{lee2016effects}.
Further work for understanding the non-linear process in $n=4$ RMP ELM suppression in EAST is undergoing.
Overall, these promising results expand previous physical understanding on ELM suppression window and demonstrate the potential effectiveness of RMPs for reliably controlling ELMs in the pre-fusion power operation phase of the ITER research plan.
\ack
\label{sec:acknowledgments}
This work is supported by the Natural Science Foundation of Anhui Province under Grant No. 2208085J39, the National Key R\&D Program of China under Grant No. 2017YFE0301100 and the National Natural Science Foundation of China under Grant No. 11875292.
\section*{References}
\bibliographystyle{ieeetr}
\bibliography{Refer.bib}
\end{document}